\documentstyle[preprint,aps]{revtex}
\tightenlines
\begin{document}
\title{ Random path averaging in multiple scattering theory\footnote{To appear in Physical Review B1, February 1, 97}}
\author{V. S. Podolsky and A. A. Lisyansky}
\address{Department of Physics, Queens College, CUNY, Flushing, NY 11367}  
\date{\today}
\maketitle

\begin{abstract}
A new method of path averaging for waves propagating in a random dilute system of identical
scatterers is developed. The scattering matrix of such a system is calculated. The method
systematically takes into account repeating scatterings on the same scatterer and effects of 
correlations. Results obtained show the significant influence of new effects on the extinction of a
coherent field  and are valid in both the diffusive and localized regimes.

\end{abstract}

\section{Introduction}
Diffusion and localization of waves propagating in random media have attracted a great deal of
interest within the last decade \cite{sheng,soucoul}. Even though considerable progress in
understanding these phenomena has been achieved, many questions remain open. A commonly
accepted picture is as follows. A coherent wave incident on a medium breaks into an incoherent
stream of photons within the distance of the order of the photon scattering mean-free path, $ \ell$ which then travel randomly through the medium. Thus, one can expect a smooth diffusive
distribution of the wave amplitude and strong fluctuations of the phase at space-time scales 
of the order of the photon mean-free path and the free-flight time. When the wavelength,
$\lambda$, is much smaller then $\ell$, the transport equation or the random walk approach give
an adequate description of the photon diffusion in a medium \cite{miln}. On the other hand, when
$\lambda$ 
and $\ell$ are comparable, wave interference can destroy ordinary diffusion and can lead to the wave localization. To incorporate interference effects into the theory of wave
propagation in random media one must start with the solution of the wave equation,
\begin{equation}
\left[\omega^2\left(1+\cal E({\bf x})\right)+\triangle\right]\Psi({\bf x})=0. 
\end{equation}
Here $\omega$ is the wave frequency and the function $\cal E({\bf x})$ describes properties of
scatterers and their spatial distribution. When self-averaging of a field takes place in a random
medium, observable quantities can be obtained upon averaging over an ensemble of random
configurations of scatterers. An averaged Green function of Eq.\ (1),
$\langle\mbox{\large$\hat{\bf G}$}_{{\bf x}}^{}\rangle$, 
defines the extinction length, $\ell_{ex}$, and the density of states. The transport
properties of waves are described by the field-field correlation function,  
$\langle\mbox {\large$\hat{\bf G}$}_{{\bf x}}^{}\mbox {\large$\hat{\bf G}$}_{{\bf
x}'}^*\rangle$.
With the help of the Dyson equation, analysis of \ 
$\langle\mbox {\large$\hat{\bf G}$}_{{\bf x}}^{}\rangle$ \ is usually reduced to calculations of
the mass-operator. The field-field correlator satisfies the Bethe-Salpiter equation. The latter
involves the vertex function which describes an effective interaction of waves in a random
medium. Microscopic analysis of these functions in a system of identical scatterers is based on
multiple scattering perturbation theory. The perturbation procedure represents all quantities of
interest as infinite series, where each term involves free-space Green functions, one-particle
scattering operators and, partial distribution functions of scatterers. Each term in the series
corresponds to a particular sequence of scattering events and has a multiplier \ $n^m$, where \
$n$ is the average density of particles and $m$ is the number of different scatterers participating
in the scattering process. The theory contains three fundamental length scales, namely, the
scatterer size, $a$, the separation between scatterers, $n^{-1/3}$, and
the wavelength, $\lambda$. In the case of dilute systems, when the volume fraction of
scatterers, $na^{3}$, is small, one can ignore all correlations in a scatterers distribution and use
only the first order terms of the mass-operator and the vertex function as the leading terms of the
expansions over small $n$.  Most of the analytical results are based on this {\it 
leading approximation} in the low density limit and on simple corrections to it.

The Green function and field-field correlator obtained in the leading approximation only takes a a restricted class of scattering diagrams into account. Only multiple scattering processes in
which a wave scatters no more than once off any scatterer are included.  This
prescribes a ``long memory'' to a wave, such that when traveling in a medium it ``remembers'' all
scatterers that were already visited and avoids them in its subsequent motion. In
application to the wave diffusion, this restriction seems to contradict to the underlying picture of
randomly walking photons. In the case of  particle-like photons this would lead to anomalous
diffusion, significantly different from ordinary diffusion \cite{ziman,andiff}. Besides, the
fraction of these ``long memory'' paths is exponentially small for long path lengths. Therefore, the
leading approximation must be accompanied by strong physical reasons for ignoring other statistically dominating scattering sequences. Such reasoning usually appeals to destructive
interference of scattered waves which separates wave diffusion from diffusion of particles. This
was analyzed in numerical calculations on finite clusters of scatterers \cite{numcal}, but reliable
analytical results in this area are absent. The problem arises because, in order to investigate an
influence of wave interference, one must evaluate the entire series for the Green function and the
field-field correlator. 

In this paper we propose a new method that enables us to consider processes with repeated
scattering in calculation of the scattering operator. Our method is based upon ({\em i}) averaging
over random paths instead of averaging over ensemble configurations, ({\em ii}) the statistical
independence of random jumps of scattered waves, ({\em iii}) the existence of a hierarchy of
length scales in dilute systems of scatterers. 

\section{Low density leading approximation}
The Green function of Eq.\ (1) corresponds to a field induced by a $\delta$--source. 
It satisfies an integral equation with the iteration solution given by the formal operator series:

\begin{equation}
\mbox {\large$\hat{\bf G}$}=\!\mbox {\large$\hat{\bf D}$}+\!\mbox {\large$\hat{\bf
D}$}\mbox{\large$\hat{\bf\cal E}$}\mbox{\large$\hat{\bf D}$}+\!\mbox{\large$\hat{\bf
D}$}\mbox{\large$\hat{\bf\cal E}$}\mbox{\large$\hat{\bf D}$}\mbox{\large$\hat{\bf\cal E}$}
\mbox{\large$\hat{\bf D}$}\!+\ldots=\!\mbox {\large$\hat{\bf D}$}+\!\mbox {\large$\hat{\bf
D}$}\mbox{\large$\hat{\bf T}$}\mbox {\large$\hat{\bf D}$}\;,
\end{equation}
where $\large\hat{\bf D}$ is a free-space propagator and $\large\hat{\bf T}$ denotes the total scattering operator of a system. One can choose any representation to obtain a constructive form of these expressions. For
instance, in the coordinate representation, operators  $\mbox {\large$\hat{\bf D}$}$ and 
$\mbox{\large$\hat{\bf\cal E}$}$ have the following matrix elements:

\begin{equation}
{\large\bf D}_{\bf x\,x'}=\frac{\mbox{\large$e^{iq|{\bf x-x'}|}$}}{4\pi|{\bf x\!-\!x'}|}\;,\;
\hspace{.5in}{\large\bf\cal E}_{\bf x\,x'}= q^2\, {\bf\cal E({\bf x})}\,\delta_{\bf x\,x'}\;, 
\end{equation}
where  $q\!=\!\omega$  is a wave number in the system of units where speed of waves  $c=1$.

In the case of a single scatterer, using an exact solution of Eq.\ (1), one can obtain the  scattering
operator of a single particle, $\mbox{\Large$\hat{\bf t}$}$. In a system of $N$ identical
scatterers, the total scattering operator can be expressed in terms of $\mbox{\Large$\hat{\bf
t}$}$--operators \cite{frish}:

\begin{equation}
\mbox {\large$\hat{\bf T}$}=\sum_{s=1}^{\infty}\;\sum^{}_{({\bf x^{}_{1}\!\ldots 
x^{}_{s}})}\:\mbox{\Large$\hat{\bf t}$}^{}_{{\bf x}^{}_{s}}\mbox {\large$\hat{\bf
D}$}\!\ldots
\mbox {\large$\hat{\bf D}$}\mbox{\Large$\hat{\bf t}$}^{}_{{\bf x}^{}_{2}}\mbox
{\large$\hat{\bf D}$}\mbox{\Large$\hat{\bf t}$}^{}_{{\bf x}^{}_{1}}\;,  
\end{equation}
where $\mbox{\Large$\hat{\bf t}$}^{}_{{\bf x}^{}_{m}}$ denotes the scattering operator of a
particle located at the point  ${\bf x}^{}_{m}$. The sums in Eq.\ (4) account for all scattering
paths, \mbox{(${\bf x^{}_{1}\!\rightarrow x^{}_{2}\rightarrow\!\ldots\!\rightarrow
x^{}_{s}}$)},
that are possible within a given spatial distribution of particles. Ensemble averaging of Eq.\ (4)
eliminates this restriction and gives an average scattering operator 
$\langle\mbox{\large$\hat{\bf T}$}\rangle_{\rm ensemble}={\bf T}$ 
in a form of the weighted sum of contributions from all scattering paths:

\begin{equation}
{\bf T}=\sum_{s=1}^{\infty}\:\int^{}_{({\bf x^{}_{1}\!\ldots x^{}_{s}})}W({\bf x^{}_{1}
\!\ldots x^{}_{s}})\mbox{\Large$\hat{\bf t}$}^{}_{\bf x^{}_{s}}\mbox
{\large$\hat{\bf D}$}\!\ldots \mbox
{\large$\hat{\bf D}$}\mbox{\Large$\hat{\bf t}$}^{}_{\bf x^{}_{2}}\mbox
{\large$\hat{\bf D}$}\mbox{\Large$\hat{\bf t}$}^{}_{\bf x^{}_{1}}, 
\end{equation}
where symbol  $\int^{}_{({\bf x^{}_{1}\!\ldots x^{}_{s}})}$ \ denotes integrations over the all
indicated variables with the natural measure for each variable. The path weight, 
$W({\bf x}^{}_{1}\!\ldots {\bf x}^{}_{s})$, is the fraction of those configurations where the
path is allowed and it can be expressed in terms of partial distribution functions of scatterers.

The conventional way of calculating \ ${\bf T}$ \ in a statistically homogeneous system makes use
of the momentum representation where the free-space propagator is diagonal,  
${\bf D}^{}_{\bf k\,k'}=\delta^{}_{\bf k\,k'}\,{\large\bf
D}_{k}^{\omega}=\delta^{}_{{\bf k\,k'}}\left[{\bf k}^2-\omega^2\right]^{-1}$, 
and the scattering matrix of a single particle  depends upon the location of a scatterer in a simple
way \   
$\langle{\bf k'}|{\mbox{\Large$\hat{\bf t}$}}_{{\bf x}}|{\bf k}\rangle=\mbox{\Large${\bf
t}$}^{}_{\bf k'\,k}\,\exp{{\em i}{\bf x(k'\!-\!k)}}$. 
Ignoring all short-range correlations arising from the rigid nature of scatterers, one can calculate
exactly the ``long memory'' subseries. This subseries contains those terms of \mbox{Eq.\ (5)}
where the scattered wave visits different scatterers only, so, all \ ${\bf x}_{m}$ are independent
and \ $W({\bf x^{}_{1}\!\ldots x^{}_{s}})=n^{s}$. Retaining only these terms is known as the
low density {\em leading approximation} or as the approximation of independent scattering. It is
thought to be appropriate in the case of dilute systems of small scatterers and most of analytical
results are based on it. For instance, the diagonal part of the ${\bf T}$--operator in this
approximation has the form \cite{sornette}:

\begin{equation}
{\bf T}_{k,\omega}^{\rm leading}=n
\mbox{\Large${\bf t}$}_{k,\omega}\left[1\!-n\mbox{\Large${\bf
t}$}_{k,\omega}{\large\bf
D}_{k}^{\omega}\right]^{-1}, 
\end{equation}
where \  
$\mbox{\Large${\bf t}$}_{k,\omega}=\langle{\bf k}|\mbox{\Large$\hat{\bf t}$}|{\bf
k}\rangle$ \  
is the scattering amplitude of a single particle in the forward direction.

However, even a naive statistical analysis gives rise to doubts regarding this approximation. Let
us consider a system with the linear size \  $L$ \ containing \ $N$ \ scatterers. A wave propagates
through the system by means of random jumps from one scatterer to another. During this motion
a likely jump length is of the order of the mean-free path, $\ell$. In the diffusion limit after \ $s$ \
jumps a wave covers the distance \ $\ell\sqrt{s}$ \ and leaves the system when \ $s\geq
s_{0}=L^{2}\ell^{-2}$. Therefore, paths with length of the order of \ $s_{0}$ \ are
most responsible for the diffusive content of an emerging field. A total number of paths consisting
of \ $s$ \ jumps is equal to \  $N^s$. The number of ``long memory'' \ $s$-paths is \
$N!/{(N\!-\!s )!}$. It is straightforward to estimate that in a macroscopic system the ratio of 
these two numbers, or fraction of the ``long memory'' paths, tends to zero as \
$\exp{\left[\,-\,{\sqrt{s}}/{2n\sqrt[3]{\ell}}\,\right]}$. Thus, all ``long memory'' paths represent a
statistically negligible fraction of the scattering paths that are important for the wave diffusion. 

Moreover, the ``long memory'' in random walks provides an effect of ``outward pressure'' acting
on a walker. For a simple walker the probability to stay within any volume around the origin after
it makes \ $s$ \ steps is proportional to \ $N^s/\,V^s\,=n^s$. When a walker is forced to avoid
repeating scattering on the same scatterer, this probability is proportional to \
$N!/\,\left(N-s\right)!s! V^s\,\approx n^s/s!$. The longer paths we consider the smaller,
comparing with an unrestricted walker, this probability is. In the latter case an average distance from the origin, $R_s$, is of the order of \ $\ell\sqrt{s}$. Since for a particle with the ``long memory'' the probability to stay within any volume around the origin is weakened by the factor \ $1/s!$, the \ $R_s$ must be greater than that for a simple walker. This can be treated as an
effective outward pressure leading a particle much farther away from the origin. The similar
pressure, caused by excluded volume, exists in the case of the self-avoiding random 
walks and it leads to anomalous diffusion \cite{ziman,andiff}.

From this point of view the success of the leading approximation in a description of the wave
diffusion looks rather surprising. In any case, it is necessary to investigate the effect of all
remaining paths on the scattering operator and the field-field correlator.

\section{Random path averaging}

With the use of uncorrelated distributions of scatterers one can  calculate the ``long memory''
subseries exactly. However, it does not help in the evaluation of the total scattering operator
without bias against any scattering paths. To advance here we modify the averaging procedure.
Averaging over random paths of scattered waves is physically equivalent to averaging over
random distributions of scatterers. Therefore, we can rewrite Eq.\ (5) using jump vectors, 
${\bf R}^{}_{m}\!=\!{\bf x}^{}_{m+1}\!\!-\!{\bf x}^{}_{m}$, as
independent variables instead of coordinates of scatterers 

\begin{equation}
{\bf T}\!=\!\sum_{s=0}^{\infty}\,\int_{({\bf x}^{}_0,{\bf R}^{}_{1},\ldots,{\bf R}^{}_{s})}^{}\!\overline{W}\,
\mbox{\Large$\hat{\bf t}$}^{}_{{\bf x}^{}_0\!+\!{\bf R}^{}_{1}+\ldots+{\bf R}^{}_{s}}
\mbox {\large$\hat{\bf D}$}\!\ldots \mbox {\large$\hat{\bf D}$}\mbox{\Large$\hat{\bf
t}$}^{}_{{\bf x}^{}_0\!+\!{\bf R}^{}_{1}+\!{\bf R}^{}_{2}} \mbox {\large$\hat{\bf D}$}\mbox{\Large$\hat{\bf
t}$}^{}_{{\bf x}^{}_0\!+\!{\bf R}^{}_{1}}\!
\mbox {\large$\hat{\bf D}$}\mbox{\Large$\hat{\bf t}$}^{}_{{\bf x}^{}_0}\;. 
\end{equation}
Any possible scattering path, $\{{\bf x}^{}_0,{\bf R}^{}_{1}\!\ldots\!{\bf R}^{}_{s}\}$, is accounted in Eq.\ (7) with a path weight, $\overline{W}({\bf x}^{}_0,{\bf R}^{}_{1}\!\ldots\!{\bf R}^{}_{s})$. In the system of $N$ scatterers fixed in space, the total number of \ $s$-paths is equal to  \ $\left(N\!-\!1\right)^{s}$ \ and, as one can see from Eq.\ (4), they are all accounted for in \mbox{\large$\hat{\bf T}$}. When scatterers are randomly placed in space, a path weight is proportional to the probability to find all scatterers in proper places along a path. This probability is given by a joint distribution function of \ $s$ \ scatterers which takes into account all correlations between them. If one ignores these correlations, a path weight takes a form used in the leading approximation, $\overline{W}\!=\!\left(\frac{N\!-\!1}{V}\right)^{s}\!=\!n^{s}$. Any attempt to preserve all correlations in \ $\overline{W}$ \ presents too general problem with no constructive way to go far beyond the leading approximation. 
Meanwhile, a random path picture allows us to approach the problem utilizing the {\it random jump distribution} instead of distribution functions of scatterers. On each random step along the path   $\{{\bf x}^{}_0,{\bf R}^{}_{1}\!\ldots\!{\bf R}^{}_{s}\}$  a wave may jump to any of \ $N\!-\!1$ \ scatterers. If we introduce a density of scatterers available for a single jump, $ng(R)$, and assume statistical independence of random steps, then the path weight can be factorized as follows:
\begin{equation}
\overline W({\bf x}^{}_0,{\bf R}^{}_{{}_1}\!\ldots {\bf R}^{}_{s})\!=n^{s+1}
g(R^{}_{{}_1})\!\ldots
\!g(R^{}_{s})\,.
\end{equation}
The function  $ng(r)$ discribes the density of scatterers at a distance $r$ from the one placed at the origin. In a statistically homogeneous system this function is close to the average density, $n$. However, correlations between scatterers and fluctuations of their concentration cause deviations of $g(r)$ from unity at small distances. This is supported by the following argument.  The fluctuations of density are discribed by the Poisson distribution. In a dilute system an average volume per particle, $n^{-1}$, is much greater than the volume of a scatterer, $a^3$. Therefore, analyzing density fluctuations at large scales we can neglect the volume of the scatterers themself. The probability to find \ $m$ \ particles within a spherical shell of radius \ $r$ \ and thickness \ $n^{-1/3}$ is \  
$P(m,r)\!=\!(4\pi r^2n^{\frac{2}{3}})^{m}\exp{(-4\pi r^2n^{\frac{2}{3}})}/m!$. It shows that the probability of large fluctuations of density rapidly decreases at \ $r\gg n^{-1/3}$. We can expect, therefore, that the deviations of \ $g(r)$ \ from unity are substantial at distanses comparable to the average separation between particles, $n^{-1/3}$. In application to wave propagation further simplification seems to be relevant. When the scattering crossection, $\sigma_{\rm sc}$, is small the precise structure of \ $ng(r)$ \ more likely can be disregarded since a wave with a large probability can just ignore a lot of surrounding scatterers. The probability that a wave makes an ${\bf r}$--jump is proportional to the probability of not meetting any scatterer inside a cylinder of lenght \ $r$ \ with the base area equal to $\sigma_{\rm sc}$. From the Poisson distribution it follows that 
\ $P(0,r\sigma_{\rm sc})\!=\!\exp{(-nr\sigma_{\rm sc})}$. Therefore,  
we can assume that the effect of density fluctuations on \ $ng(r)$ \ is substantial at distances of the order of the scattering mean-free path, $(n\sigma_{\rm sc})^{-1}$. The function that takes into account the above arguments can be choosen in the form
\begin{equation}
ng(r)=n\left[1-\frac{\gamma^{3}}{8\pi n}\exp(-r\gamma)\right]. 
\end{equation}
This function is normalized on $N\!-\!1$ and, consequently, the path weight $\overline W({\bf
x}^{}_0,{\bf R}^{}_{{}_1}\!\ldots {\bf R}^{}_{s})$ is normalized on the total number of $s$-paths. 
A parameter \ $\gamma$ \ is understood to be of the order of \ $n\sigma_{\rm sc}$, if the
scattering mean-free path is much greater than \ $n^{-1/3}$, and it tends to the inverse average separation, $n^{-1/3}$, when the scattering cross section becomes large and both lengths become  comparable. 

In a dilute system Eq.\ (7) can be further simplified by noting that the size of a scatterer is
much smaller than the average separation. If we use the coordinate representation in all
intermediate states for the matrix elements 
$\langle{\bf k'}|\mbox {\large$\hat{\bf T}$}|{\bf k}\rangle={\bf T}^{}_{\bf k' k}$,

\begin{equation}
{\bf T^{}_{k' k}}\!=
\!\sum_{s=0}^{\infty}\:
\int_{({\bf x}^{}_0,{\bf R}_1\ldots{\bf R}_s)}^{}\!
{\overline W}
\mbox{\large$e^{i{\bf k'}({\bf x}^{}_0\!+\!{\bf R}^{}_{s}\!+\!\ldots\!+{\bf R}^{}_{1})-i{\bf kx}^{}_0}$}
\int_{({\bf r}^{}_1,{\bf r}_{1}' \ldots {\bf r}^{}_s,{\bf r}_{s}')}^{}\!
\mbox{\Large${\bf t}$}^{}_{{\bf k'}{\bf r}_{s}'}
{\bf D}_{{\bf r}_{s}'-{\bf r}_{s}+\!{\bf R}^{}_{s}}^{}
\cdots
{\bf D}_{{\bf r}_{1}'-{\bf r}^{}_{1}+{\bf R}^{}_{1}}^{}
\!\mbox{\Large${\bf t}$}^{}_{{\bf r}^{}_{1}{\bf k}}\:,
\end{equation}
we can see that the ranges of changes of variables $({\bf R}_{1}\ldots{\bf R}_{s})$ and
 $({\bf r}^{}_1,{\bf r}_{1}' \ldots {\bf r}^{}_s,{\bf r}_{s}')$ \ have
different order of magnitude. A single particle scattering matrix in the coordinate representation,
$\mbox{\Large${\bf t}$}_{{\bf r}'{\bf r}}$, vanishes outside the scatterer, 
where $r'\!,r\geq a$, while the distances between two successive scatterers, $R$, are of the order
of the average separation, $n^{-1/3}$. Assuming that short-range density fluctuations are not
crucially important in the low density limit, we can assume that essential values of \ $R$ \ are much
greater than those of \ $r',r$. This fact allows one to use  
the far-zone asymptote of the free-space propagators in \mbox{Eq.\ (10)}:

\begin{equation}
{\bf D}^{}_{\bf R\!+\!r\!-\!r'}\approx
{\bf D}^{}_{R}\mbox{\large$e$}^{i{\bf q(r\!-\!r')}},\hspace{.3in} 
{\large\bf D}_R=\frac{\mbox{\large$e^{iqR}$}}{4\pi R}\;,
\end{equation}
where ${\bf q}=\mbox{\small$\omega\hat{\bf n}$},\:\mbox{\small$\hat{\bf n}$}={\bf R}/R.$
It makes it possible to calculate all integrals with respect to $({\bf r}^{}_1,{\bf r}_{1}' \ldots {\bf r}^{}_s,{\bf r}_{s}')$ in Eq.\ (10):
\begin{equation}
{\bf T}^{}_{\bf k'k}\!\approx\!
\sum_{s=0}^{\infty}
\int^{}_{{}_{({\bf x,R\!\ldots})}}\!\!
{\overline W}
\mbox{\large$e^{i{\bf k'}({\bf x}^{}_0\!+\!{\bf R}^{}_{s}\!+\!\ldots\!+{\bf R}^{}_{1})-i{\bf kx}^{}_0}$}
{\mbox{\Large${\bf t}$}^{}_{{\bf k'}{\bf q}^{}_{s}}}\!\cdots
\mbox{\Large${\bf t}$}^{}_{{\bf q}^{}_{1}{\bf k}}
{\bf D}^{}_{R^{}_{s}}\!\cdots{\bf D}^{}_{R^{}_{1}}\:.
\end{equation}

In a homogeneous system, the integral over the location of the first scatterer, ${\bf x}^{}_0$, gives the 
common multiplier \ $\delta^{}_{\bf k'k}$ \ expressing homogeneity of 
the system. Since the distribution function, $g(R)$, depends upon the jump length only, the integrations
over the directions of the jumps can be separated from the integrals over jump lengths in \mbox{Eq.\ (12)}. All
directional integrals can be evaluated with the help of the Taylor expansion of smooth
functions in the integrand and a deformation of the integration contour in the complex plane: 
\begin{equation}
\int_{\mbox{\small$\hat{\bf n}$}}^{}
{\mbox{\Large${\bf t}$}}^{}_{\!\ldots\mbox{\small$\hat{\bf n}$}}
{\mbox{\Large${\bf t}$}}^{}_{\mbox{\small$\hat{\bf n}$}\ldots}\!
\mbox{\large$e$}^{ikR\mbox{\small$\hat{\bf n}$}\mbox{\small$\hat{\bf k}$}}\!=
\int_{0}^{2\pi}\!d\phi\int_{-1}^{1}\!d\mu f(\phi,\mu)\mbox{\large$e$}^{ikR\mu}
\approx
\sum_{\sigma=\pm}\,\frac{2\pi{\scriptstyle\sigma}}{ikR}
\mbox{\Large${\bf t}$}_{\ldots\sigma\omega\mbox{\small$\hat{\bf k}$}}^{}
\mbox{\Large${\bf t}$}_{\sigma\omega\mbox{\small$\hat{\bf k}$}\ldots}^{}
{\mbox{\large $e$}^{ikR\sigma}}.
\end{equation} 
This formula works quite well in the case of weak scattering anisotropy,
\begin{equation}
\frac{\partial\mbox{\Large${\bf t}$}^{}_{\!\ldots\mbox{\small$\hat{\bf n}$}}}
{\partial(\mbox{\small$\hat{\bf n}$}\cdot\mbox{\small$\hat{\bf k}$})}\ll 
kR\;\mbox{\Large${\bf t}$}^{}_{\ldots\mbox{\small$\hat{\bf n}$}}\:,
\end{equation} 
especially when \ $kR\gg 1$. Eq.\ (13) becomes the rigorous equality for isotropic scattering. After
utilizing Eq.\ (13), all the remaining integrals over jump lengths in Eq.\ (12) can be factorized since there is no correlations beween jumps. The resulting expression for diagonal elements of ${\bf
T}^{}_{\bf k'k}$ takes the form
\begin{equation}
{\bf T}^{}_{k\,\omega}=\!
\!n\mbox{\Large$\bf t$}^{}_{\bf k\,k}\!+\!
n\sum_{s=1}^{\infty}\sum_{\sigma^{}_{1}\sigma^{}_{2}\cdots}^{}\!
{\bf t}^{}_{{\bf k}\sigma^{}_{s}}
g_{\sigma^{}_{s}}
{\bf t}^{}_{\sigma^{}_{s}\sigma^{}_{s\!-1}}
\ldots
g_{\sigma^{}_{2}}
{\bf t}^{}_{\sigma^{}_{2}\sigma^{}_{1}}
g_{\sigma^{}_{1}}
{\bf t}^{}_{\sigma^{}_{1}{\bf k}}\:.
\end{equation}
In this equation the indexes $\sigma^{}_{1},\sigma^{}_{2},\cdots$ correspond to ``+'' or ``--'' and
we denote
$${\bf t}^{}_{{\bf k}\sigma}\!=\!
\langle{\bf k}|\mbox{\Large$\bf t$}|\sigma\omega\mbox{\small$\hat{\bf k}$}\rangle,\;\;
{\bf t}^{}_{\sigma{\bf k}}\!=\!
\langle\sigma\omega\mbox{\small$\hat{\bf k}$}|\mbox{\Large$\bf t$}|{\bf k}\rangle\;\;,$$
\begin{equation}
{\bf t}^{}_{\sigma\sigma'}\!=\!
\langle\sigma\omega\mbox{\small$\hat{\bf k}$}|
\mbox{\Large$\bf t$}|
\sigma'\omega\mbox{\small$\hat{\bf k}$}\rangle,\:
g_{\sigma}\!=\!\frac{{\scriptstyle\sigma}2\pi n}{ik}
\int_{0}^{\infty}
rg(r){\bf D}_{r}^{\omega}{\mbox{\large $e$}^{ikR\sigma}}dr\:.
\end{equation}
Due to the fact that all intermediate states in Eq.\ (15) are aligned along or against the  incident
momentum ${\bf k}$, the multiple-scattering part of the $\bf T$--matrix can be rewritten in terms
of $2\times 2$--matrices ${\bf t}$ and ${\bf g}$ defined in Eq.\ (16):
\begin{equation}
{\bf T}^{\rm multiple}_{k\,\omega}=\!
n\sum_{s=0}^{\infty}\sum_{\sigma\sigma'}^{}\!
{\bf t}^{}_{{\bf k}\sigma}
\left[{\bf g}\left({\bf tg}\right)^{s}\right]^{}_{\sigma\sigma'}
{\bf t}^{}_{\sigma'{\bf k}}\!=
n\sum_{s=0}^{\infty}\sum_{\sigma\sigma'}^{}\!
{\bf t}^{}_{{\bf k}\sigma}
\left[{\bf g}
\left(1\!-\!{\bf tg}\right)^{-1}\right]^{}_{\sigma\sigma'}
{\bf t}^{}_{\sigma'{\bf k}}\:.
\end{equation}

Finally, using an explicit expression for the matrix $\left(1\!-\!{\bf tg}\right)^{-1}$, we can
represent the $\bf T$--matrix in the form
\begin{equation}
{\bf T}\!=\!
n\mbox{\Large$\bf t$}^{}_{\bf k\,k}\!+\!
\frac{n}{1\!-\!{\rm tr}\,\|{\bf tg}\|\!+\!{\rm det}\,\|{\bf tg}\|}\!
\left(
{\bf t}^{}_{\scriptscriptstyle+{\bf k}},\,
{\bf t}^{}_{\scriptscriptstyle-{\bf k}}
\right)\!\!
\left(\!
\begin{array}{cc}
{1\!-\!{\bf t}^{}_{\scriptscriptstyle--}g_{\scriptscriptstyle-}\,,}&\!\!
{{\bf t}^{}_{\scriptscriptstyle-+}g_{\scriptscriptstyle+}}\\
{{\bf t}^{}_{\scriptscriptstyle+-}g_{\scriptscriptstyle-},\:}&
{1\!-\!{\bf t}^{}_{\scriptscriptstyle++}g_{\scriptscriptstyle+}}
\end{array}
\right)\!\!
\left(
\begin{array}{c}
{g_{\scriptscriptstyle+}{\bf t}^{}_{{\bf k}\scriptscriptstyle+}}\\
{g_{\scriptscriptstyle-}{\bf t}^{}_{{\bf k}\scriptscriptstyle-}}
\end{array}
\right),
\end{equation}
where  ${\rm tr}\,\|{\bf tg}\|$ and  ${\rm det}\,\|{\bf tg}\|\,$ denote the trace and the determinant,
respectively. For a spherical scatterer \ ${\bf t}^{}_{\scriptscriptstyle++}\!={\bf
t}^{}_{\scriptscriptstyle--}$ \ is the forward scattering amplitude and \  
${\bf t}^{}_{\scriptstyle-+}\!={\bf t}^{}_{\scriptscriptstyle+-}$ \ is the backscattering amplitude
with all momenta confined at the ``mass shell'', \ 
${\bf k}^{2}\!=\!\omega^{2}$, while in \ ${\bf t}^{}_{{\bf k}\pm}$ \ and 
${\bf t}^{}_{\pm{\bf k}}$ \ one of the momenta $(\bf k)$ can be of an arbitrary magnitude.

This $\bf T$--matrix is formally defined at all spatial scales and for any wavelength. In our
evaluation, we assumed a weak anisotropy of scattering and a low density of particles, 
\ $na^{3}\ll 1$. An isotropic scattering dominates, for
instance, in the Rayleigh limit when a wavelength is much greater than the scatterer size,
$a\,\omega\ll 1$. In this case, since scattering amplitudes do not depend upon directions of incoming
and outgoing momenta, Eq.\ (18) can be simplified as follows:

\begin{equation}
{\bf T}^{}_{k\,\omega}=n
\left[\mbox{\Large${\bf t}$}^{}_{k\,\omega}+ \frac{n{\cal
T}^{2}_{k\,\omega}
(g_{+}+g_{-})}{1-\mbox{\Large${\bf t}$}^{}_{\omega}(g_{+}+g_{-})}\right], 
\end{equation}
where \ 
$\mbox{\Large${\bf t}$}^{}_{\omega}=\langle\omega\mbox{\small$\hat{\bf
k}$}|\mbox{\Large$\hat{\bf t}$}|\omega\mbox{\small$\hat{\bf k}$}\rangle$ \ 
is the on-shell scattering amplitude \ 
${\cal T}^{}_{k,\omega}=\langle{\bf k}|\mbox{\Large$\hat{\bf
t}$}|\omega\mbox{\small$\hat{\bf k}$}\rangle$ \ 
is the on-off-shell amplitude and \ 
$\mbox{\Large${\bf t}$}^{}_{k,\omega}=\langle{\bf k}|\mbox{\Large$\hat{\bf
t}$}|{\bf k}\rangle$ \ 
is the off-shell one. All these quantities can be obtained from the solution of the
boundary-value problem for a single scatterer and are well-known for scatterers of a simple shape
\cite{morse,kirkpat,single}. Quantities \ $g_{\pm}$ are given in Eq.\ (16) and depend upon the choice
of the density of surrounding scatterers. The substitution of the function $g(r)$ given by Eq.\ (9)
yields the total scattering operator of a dilute system of identical isotropic scatterers: 

\begin{equation}
{\bf T}^{}_{k\,\omega}\!=n
\frac{\mbox{\Large${\bf t}$}^{}_{k\,\omega}\!+
n\!
\left[
{\cal T}^{2}_{k\,\omega}\!-
\mbox{\Large${\bf t}$}^{}_{k\,\omega}\mbox{\Large${\bf t}$}^{}_{\omega}
\right]
\!\left[{\large\bf D}_{k}^{\omega}\!-
\frac{\gamma^{3}}{8\pi n}{\large\bf D}_{k}^{\omega+i\gamma}\right]}
{1-n\mbox{\Large${\bf t}$}^{}_{\omega}
\left[{\large\bf D}_{k}^{\omega}-
\frac{\gamma^{3}}{8\pi n}{\large\bf D}_{k}^{\omega+i\gamma}\right]}.
\end{equation}

This result involves the effective correlation radius $\gamma^{-1}$. In the limit \
$\gamma\rightarrow\infty$, when correlations between scatterers can be ignored, ${\bf T}$--matrix given by Eq.\ (20)
reproduces the expression for the coherence length obtained in the leading approximation. The
choice of \ $\gamma\sim\ell^{-1}_{\rm sc}$ \ physically means that the contribution to the
coherent field from scattering paths consisting of many short jumps is assumed to be small,
since they are rare in a statistically homogeneous dilute system. 

\section{extinction length}

The averaged Green function describes the coherent part of the field created by a point source placed
in a random medium. Due to random scattering of propagating waves the pole of the Green
function is shifted away from the pole of the free-space propagator, $k\!=\!\omega$. The imaginary
part of this pole defines the extinction length of coherent waves in the random medium. According
to the relationship between \ 
${\bf G}_{k,\omega}$ \ and the scattering matrix, \ 
\mbox{${\bf G}_{k\,\omega}\!=\!{\bf D}_{k}^{\omega}\!+\!\left({\bf
D}_{k}^{\omega}\right)^2
{\bf T}_{k\,\omega}\:,$} 
it seems that \ ${\bf G}_{k\,\omega}$ \ has a pole of the second order when \ $k$ \ tends to \
$\omega$. However, this singularity  disappears, since \ ${\bf T}_{k\,\omega}$ \ has there a zero
of the first order, \mbox{${\bf T}^{}_{k\,\omega}\!=\!-2\omega( \!k\!-\!\omega)\!+
\!{\cal O}\left((\!k\!-\!\omega)^2\right)$.} Our result, as well as \ ${\bf
T}_{k\,\omega}^{\rm leading}$, demonstrates this behavior. Therefore, only a pole in ${\bf
T}$--matrix gives an actual singularity of the Green function. In the leading approximation this
pole can be found from the equation:

\begin{equation}
k^{2}-\omega^{2}=n\mbox{\Large${\bf t}$}^{}_{k,\omega}
\end{equation}
This approximation relies on the low density of particles and weak scattering, which implies that \
$f\!=\!\frac{4\pi}{3}a^{3}n\ll 1$ \ and \ $\omega\mbox{\Large${\bf t}$}^{}_{k,\omega}$ \ is 
small. Weak scattering takes place in the Rayleigh limit where \ $a\,\omega\ll 1$ 
\ and the scattering amplitude itself is a small quantity. An expansion over the small parameter 
$a\,\omega$  yields the leading term of the scattering amplitude 

\begin{equation}
{\rm Re}\,\mbox{\Large${\bf t}$}^{}_{k,\omega}\approx
\varepsilon \frac{4}{3}\pi a^{3}\omega^{2},
\end{equation}
where \ $\varepsilon$ \ is the difference between dielectric constants of the scatterer and the host
medium. The ${\rm Re}\,\mbox{\Large${\bf t}$}^{}_{k,\omega}$ does not depend upon  \ $k$ \
and is of the second order, while the leading term of the imaginary part is of the fifth order, as it
follows from the optical theorem, 
$${\rm Im}\,\mbox{\Large${\bf t}$}^{}_{k,\omega}\!= \left(2\pi^2\right)^{2}\omega\int\left|
\mbox{\Large${\bf t}$}^{}_{\bf k\,k'}\right|^{2}d\mbox{\small$\hat{\bf k'}$}.$$
Solving Eq.\ (21) for long waves in a dilute system we obtain
$$k^{2}(\omega)=\omega\left[1+f\varepsilon+i(f\varepsilon)^2\frac{\omega^{3}}
{4\pi n}\right].$$
When $f\varepsilon\ll 1$, this equation gives the well-known expression for the extinction
length,  
$$l_{ex}^{-1}\!=\!{\rm Im}\,k(\omega)\!=\! n\sigma_{sc}/2\!\approx
\!\omega(f\varepsilon)^2\omega^3/{8\pi n}.$$ 
Here the factor $\omega^3/{8\pi n}$ is of the order of \ $(R/\lambda)^{3}$, where $R$ and
$\lambda$ are the average separation between scatterers and the wave length, respectively. It can
be dangerously large, but in the weak scattering regime it is totally neutralized by the small
parameter \ $f\varepsilon$. However, even in a rare mist of water drops this parameter can be of
the order of unity due to the large value of $\varepsilon$. In similar dilute systems with large
mismatch indexes the leading approximation still seems to be valid. For $R\gg\lambda$ it gives   
$$l_{ex}^{-1}\sim\omega f\varepsilon\left(R/\lambda\right)^
{\frac{3}{2}} \gg \omega.$$
On the other hand, the extinction length (the phase coherence length) being much smaller than the
wave length is physically meaningless since then we cannot talk about wave propagation anymore.
Therefore, the leading approximation requires that at least two condition to be met, namely,
$a\omega\ll 1$ \ and \ $ f\varepsilon\ll 1$.

Our ${\bf T}$--matrix was obtained under the conditions \ $f\ll 1,\; a\omega\ll 1$ and it involves a
phenomenological parameter, $\gamma\!=\!\ell^{-1}_{\rm corr}$. Our equation for the pole of the
Green function reads:
\begin{equation}
\left[k^2\!-\!\omega^{2}\right]\left[k^2\!-\!\left(\omega\!+\!i\gamma\right)^{2}\right]\!=\!
n\mbox{\Large${\bf t}$}^{}_{\omega}
\left[k^2\!-\!\left(\omega\!+\!i\gamma\right)^{2}\!-\!
\frac{\gamma^{3}}{8\pi n}\left(k^2\!-\omega^2\right)\right]
\end{equation}
To be consistent, we have to use the Rayleigh scattering amplitude given by \mbox{Eq.\ (22).} 
When \ $f\varepsilon\ll 1$, the right hand part of Eq.\ (23) is small and we obtain
$$l_{ex}^{-1}\approx\gamma.$$ For \ $f\varepsilon\sim 1$ \ and \ $R\gg\lambda$,
taking into account that $\lambda/\ell_{\rm corr}\ll 1$  and  
\mbox{$\gamma^3/8\pi n\sim(R/\ell_{\rm corr})^{3}\sim 1,$} \ we have:
$$l_{ex}^{-1}\approx\omega(f\varepsilon)^2\frac{\omega^3}{8\pi n}\left(1\!-\frac{\gamma^{3}}{8\pi n}\right).$$
In both cases we obtain physically meaningful expressions for the extinction length since we have
assumed that \ $\gamma\!=\!\ell^{-1}_{\rm corr}$ approaches \ $n\sigma_{sc}$ in the weak
scattering regime and it becomes proportional to $n^{1/3}$ when the scattering cross section
becomes large.

\section{conclusions}

In summary, we developed a new method for studying interference of multiple scattered waves
that takes into account multiple returns to the same scatterer. We evaluated the scattering matrix
of a dilute system of identical particles. The ${\bf T}$--matrix obtained in \mbox{Eq.\ (20)} is
valid in the Rayleigh limit, $\omega a\ll 1$. We did not impose any other restrictions on the wave
length, therefore, our result can be used to analyze the localized regime,  $\omega\ell\sim
1$. All scattering paths and finite range of correlations are accounted for in our
approach. This results in the momentum  cut-off that appears in \mbox{Eq.\ (20)} as a truncation
of the free-space propagator, ${\large\bf D}_{k}^{\omega}$, by the term $\frac{\gamma^3}{8\pi
n}{\large\bf D}_{k}^{\omega+i\gamma}$. It also gives an extra term in the numerator of Eq.\
(20), which is missing in ${\bf T}^{\rm leading}$. In the Rayleigh limit this term is small
because the leading terms of all scattering amplitudes are equal. However, the change of the denominator is
physically important for the consistent description of the wave extinction in a system with large
mismatch indexes. Using a microscopic expression for the density function, $g(r)$, in Eq.\ (19) one can obtain an exact expression for the extinction length. According to the definition of \
$g(r)$,  this function must have the structure \ $g(r)=1-``{\rm decay\: function}''$. It guarantees the
appearance of the momentum cut-off in the denominator of the exact \mbox{${\bf T}$--matrix.} This effect is due to scattering paths with multiple returns. Even if each such path gives a negligible contribution, they altogether present a dominant amount of scattering paths and they cannot be neglected in the coherent field. 
Referring to the motivation of our work given in the introduction, we can conclude that the
interference cancellation of non-``long memory'' contributions in the coherent field is
overestimated in the leading approximation.
 
The ${\bf T}$-matrix itself does not allow one to analyze the influence of  ``long memory'' 
effects on transport properties of waves. Calculations of the field-field correlator are currently in
progress. Our method enables us to proceed without utilizing the Bethe-Salpeter equation and
does not require any additional approximations.

We wish to thank E. Kogan for the useful discussion and A. Z. Genack for reading and
commenting on the manuscript.  This work was supported by the NSF under grant No.
DMR-9632789, by the CUNY collobarative grant and by PSC-CUNY research awards.

\end{document}